# Global track finding based on the Hough transform in the STCF detector

Hang Zhou[a,b], Kexin Sun[c], Zhenna Lu[d], Hao Li[e], Xiaocong Ai[e], Jin Zhang[d], Xingtao Huang[c] and Jianbei Liu[a,b,*]

[a]State Key Laboratory of Particle Detection and Electronics, University of Science and Technology of China, Hefei 230026, China
[b]Department of Modern Physics, University of Science and Technology of China, Hefei 230026, China
[c]Key Laboratory of Particle Physics and Particle Irradiation (MOE), Institute of Frontier and Interdisciplinary Science, Shandong University, Qingdao, Shandong, 266327, China
[d]School of Science, Shenzhen Campus of Sun Yat-sen University, Shenzhen 518107, China
[e]School of Physics and Microelectronics, Zhengzhou University, Zhengzhou, Henan, 450001, China

* Corresponding author, e-mail: liujianb@ustc.edu.cn

**Abstract:** The proposed Super Tau-Charm Facility (STCF) is an electron-positron collider designed to operate in a center-of-mass energy range from 2 to 7 GeV. It provides a unique platform for physics research in the tau-charm energy region. To fulfill the physics goals of STCF, high tracking efficiency and good momentum resolution is required for charged particles with momenta from 50 MeV/c to 3.5 GeV/c. A global track finding algorithm based on Hough transform has been developed and implemented in the STCF software framework to meet this requirement. The design of the algorithm and its performance with simulation are presented in this paper.

**Keywords:** STCF, Drift chamber, Track reconstruction

## 1. Introduction

The STCF [1] is a symmetric electron-positron collider proposed to be constructed in China. It has a center-mass-energy ($\sqrt{s}$) range from 2 to 7 GeV with a peak luminosity above $0.5 \times 10^{35}$ cm$^{-2}$s$^{-1}$ at $\sqrt{s}$ = 4 GeV. The STCF is a platform for a wide range of physics studies, including tests of QCD and electroweak interactions, study of hadron spectroscopy, and searching for new physics. The baseline STCF detector consists of a tracking system, a particle identification (PID) system utilizing two different Cherenkov detector technologies in barrel and end-cap, an electromagnetic calorimeter (EMC), and a muon detector (MUD).

The charged particles to be measured by the STCT detector have momenta up to 3.5 GeV/c, while most of them have momenta less than 1 GeV/c, and a substantial fraction of them have momenta even less than 0.4 GeV/c. The ability to detect charged particles with high tracking efficiency and good momentum resolution in a large range of momentum down to 50 MeV/c is essential for the STCF detector. The tracking system in the STCF detector includes a main drift chamber (MDC) as an outer tracker and an inner tracker (ITK). Two detector options have been proposed for the ITK [1], with one being the micro-resistive well (μRWELL) detector and the other a silicon pixel detector using the CMOS monolithic active pixel sensor (MAPS) technology. It is necessary that high-performance track reconstruction algorithms be developed for the STCF tracking system.

The Hough transform is a widely used method in image processing and computer vision for detecting geometric shapes, particularly lines and circles. It has been used as a global track finding method in offline software or at the high-level trigger level for many high-energy physics (HEP) experiments, such as the BelleII experiment [2] and BESIII experiment [3]. A Hough transform-based track finding method can handle all available hits simultaneously, and therefore is insensitive to local hit inefficiency. A track finding algorithm based on the Hough transform method was developed and implemented in the STCF offline software framework (OSCAR) [4] for the STCF tracking system with the µRWELL ITK to achieve the desired track finding efficiency and resolutions of reconstructed track parameters. This algorithm finds track candidates using hits from ITK and MDC combined. This paper describes the design and optimization of the track finding algorithm and presents its performance evaluated with simulation.

## 2. The STCF tracking detectors and their simulation

The µRWELL ITK[5] comprises three independent µRWELL layers placed at radii of 6 cm, 11 cm, 16 cm, respectively. XV readout strips are used for he each of three µRWELL layers , allowing signals generated by a charged particle in a ITK layer to be reconstructed into a space point. According to previous studies, the spatial resolution of the uRWELL ITK is 100 µm in the rφ plane and 400 µm in the beamline direction, respectively.

The MDC serve as the main tracker in the STCF detector. It has 48 layers of approximately square cells at radii ranging from 200 mm to 840 mm, arranged in 8 super-layers. The super-layers are classified as either axial (A) super-layer or stereo (U,V) super-layers, and each super-layer contains 6 layers of cells. Stereo wires have an either positive or negative inclination angle to the beamline (z). Table. 1 lists some key design parameters of MDC. The spatial resolution of the MDC is expected to be better than 120 µm on average.

Table. 1 The key parameters of the MDC.

| Super-layer type | Start radius(mm) | Cells per layer | Stereo angle(mrad) |
| --- | --- | --- | --- |
| A | 200.0 | 128 | 0 |
| U | 271.6 | 160 | 39.3 to 47.6 |
| V | 342.2 | 192 | −41.2 to −48.4 |
| A | 419.2 | 224 | 0 |
| U | 499.8 | 256 | 50.0 to 56.4 |
| V | 578.1 | 288 | −51.3 to −57.2 |
| A | 622.0 | 320 | 0 |
| A | 744.0 | 352 | 0 |

The entire tracking system operates in a 1T magnetic field along the $z$-direction, and covers the polar angles from 20° to 160°. When a track has a large dip angle, it exits the MDC through its endcap, resulting in the number of layers crossed being fewer than the total number of layers in the detector. Fig. 1 shows the layout of the tracking system.

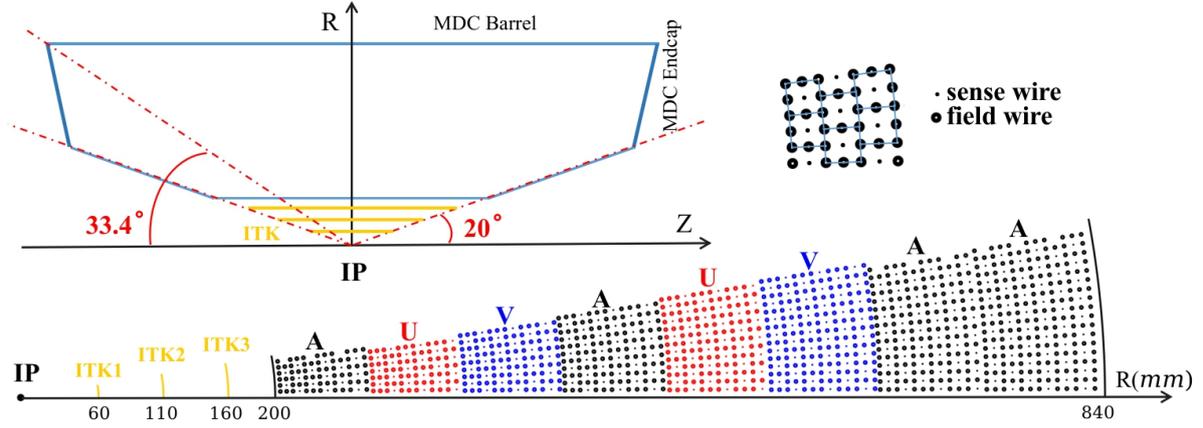

Fig. 1 The layout of the tracking system and the structure of the MDC cells.

Simulation of the STCF tracking system was performed in the OSCAR, which consists of interface to external third-party software and software packages for detector description, event generation, simulation, reconstruction and data analysis. The geometry of the entire STCF detector including the beam pipe was constructed in OSCAR. Geant4 [6] was used to produce tracks and create hits in detectors. Each hit was smeared with the expected spatial resolution of the tracking detectors and the smeared hits were used as inputs for track reconstruction.

## 3. Track finding algorithm based on the Hough transform

The main steps of track reconstruction based on Hough transform at STCF are shown in Fig. 2. Since the MDC measurements cannot directly provide the position along the beam direction, track finding starts with the searching for two-dimensional (2D) circular tracks. The MDC stereo hits are incorporated to the circular tracks afterwards to determine the three-dimensional (3D) trajectories.

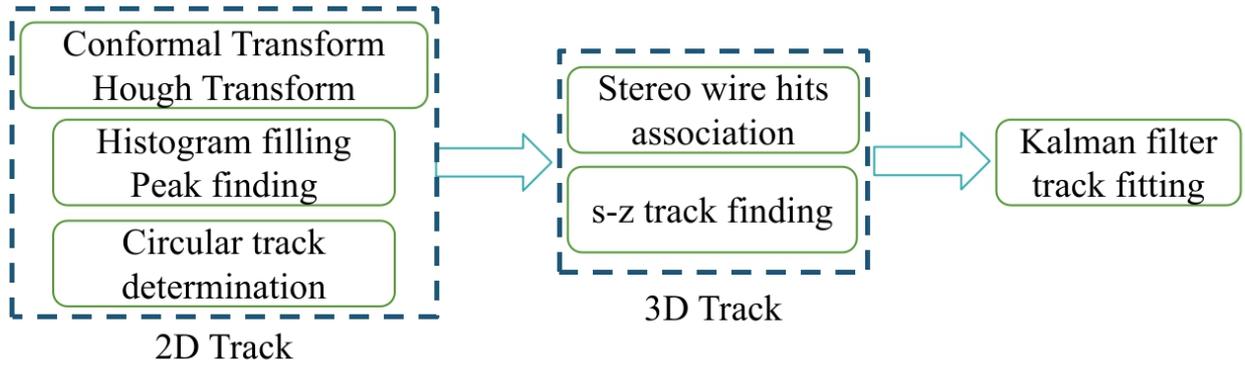

Fig. 2 The workflow of track reconstruction.

### 3.1 Track parametrization

Disregarding the effect of the materials traversed, the trajectory of a charged particle in a uniform magnetic field follows a helical path, where the projection onto the plane perpendicular to the magnetic field forms a circle. The track helix is locally represented by five parameters

$$(d_0, \varphi_0, \kappa, z_0, tan\lambda),$$

as depicted in Fig. 3. $d_0$ is the signed distance from the point of closest approach (POCA) to the $z$-axis, $\varphi_0$ is the angle of the momentum direction at the POCA, $\kappa$ is the product of the charge and the track curvature, $z_0$ is the $z$ coordinate at the POCA, $\tan\lambda$ is the tangent of the track dip angle. The projection of the helix onto the $s$-$z$ space is a straight line, where $s$ is defined as the length of the path projected in the $x$-$y$ plane.

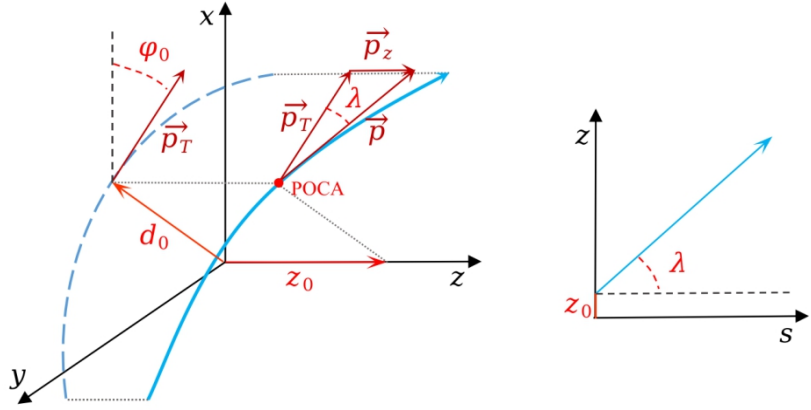

Fig. 3  Illustration of the helix parametrization of a particle track. The blue solid line represents the track. The blue dashed line in the left figure is the projection of the track onto the XY plane.

### 3.2 Conformal mapping and Hough transform

The first stage of the algorithm is to find the circular track in the transverse $x$-$y$ plane. The two-dimensional track finding starts with a conformal mapping, where each point in the geometrical x-y plane is mapped to a point($u, v$) in the conformal plane[3]. This transformation converts the circular trajectory passing through the origin into a straight line, while preserving the circular shape of the drift circle. In the conformal plane, the straight line transformed from the circular trajectory is still passing through the transformed points and tangent to those circles transformed from the drift circles. Fig. 4a and Fig. 4b shows a simulated track and its corresponding hits in both the $x$-$y$ plane and the conformal plane.

The two-dimensional track finding is thus simplified to determining the straight lines in the conformal plane. The line is described using angle-radius parameters [7] as
$$\rho = u \cdot \cos\alpha + v \cdot \sin\alpha.$$
Where $\alpha$ is the polar angle of the line's normal, and $\rho$ is its algebraic distance from the origin, as shown in Fig. 4b. This equation is used to describe the transformation of the ITK hit. The extension of the Hough transform, also known as the Legendre transform [8], is used to describe the line tangent to the drift circle in the conformal plane. Each drift circle representing a MDC measurement is mapped to a pair of sinusoids in the parameter space with the equation
$$\rho = u_w \cdot \cos\alpha + v_w \cdot \sin\alpha \pm r_{co}.$$
While ($u_w, v_w$) and $r_{co}$ represent the center and radius of the transformed drift circle in the conformal. Fig. 4c is the schematic view of mapping hits to the $\rho$-$\alpha$ space.

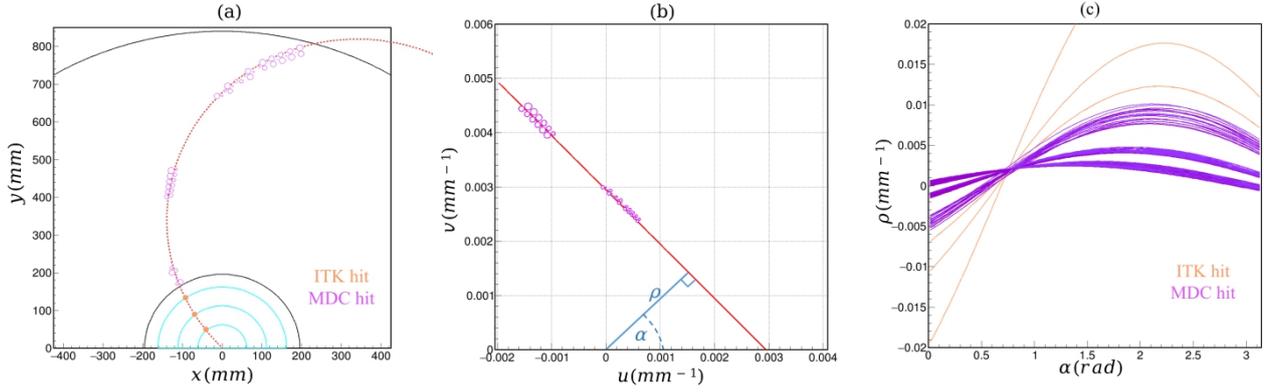

Fig. 4　The process of mapping detector measurements to Hough space.
(a) $x$-$y$ plane, (b) $u$-$v$ plane, (c) $\rho$-$\alpha$ plane.

**3.3 Two-dimensional track finding based on Hough transform**

　　Although the mapping of the ITK hit and the MDC hit to the parameter space differs to some extent, the identification of the track and the parameter determination remains consistent, i.e., locating the the region in the Hough ($\rho$-$\alpha$) space where the density of sinusoids reaches a local maximum.

**Histogram filling**

　　In the two-dimensional track finding process, hits from ITK and the axial layer of MDC are used. To find populated regions, a histogram in the $\rho$-$\alpha$ space is filled based on whether the curve passes through the bin. In the case of both pairs of sinusoids representing the MDC hit cross the same bin, the bin count increases only by one. A bin exhibiting a local maximum, where its height exceeds a predefined threshold – commonly referred to as a peak in the histogram – indicates a potential track candidate.

　　The $\rho$-$\alpha$ parameters of tracks of approximate opposite flight direction as well as opposite charges are closely aligned. In Hough space, the intersection points representing these two tracks are close to each other. Thus the peak representing one track might be contaminated by hits from another track. When searching for tracks in $\rho$-$\alpha$ space, this may result in efficiency loss of peak finding and accuracy deterioration of determined track parameters. Very low-momentum particles may generate looping tracks, and as the particles propagate, their tracks may no longer resemble helices. To reduce the negative effects of undesirable hits on track finding in the $\rho$-$\alpha$ plane, we partition each sinusoidal curve into two histograms accordingly based on whether $\rho \cdot d\rho/d\alpha$ is greater than zero or not, as shown in Fig. 5.

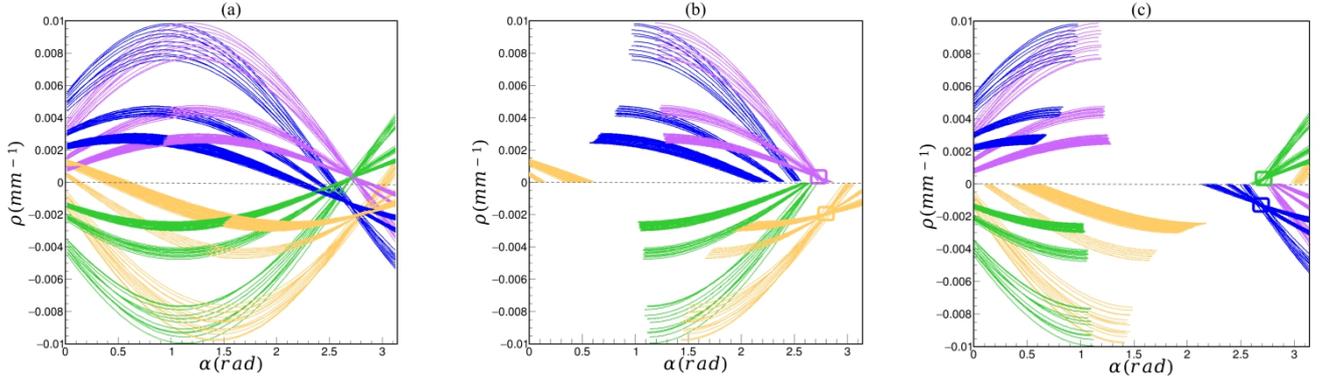

Fig. 5 An example of ψ(3686) → π+ (blue) π- (orange) J/ψ, J/ψ → μ+ (green) μ- (violet) event with 4 tracks. (a) The sinusoidal curves representing MDC hits in the $\rho$-$\alpha$ space, the sinusoidal for ITK hits are not shown.; (b) The portion of the curves where $\rho \cdot d\rho/d\alpha < 0$; (c) The portion of the curves where $\rho \cdot d\rho/d\alpha > 0$. The rectangular boxes mark the region with the maximal density of intersections.

**Optimization of bin width**

The selection of bin width is related to the errors of track and hit parameters. The errors should be included in the same bin of the histogram as much as possible, otherwise the intersection in the Hough space may be divided into two bins. On the other hand, excessively large bin widths may lead to track determination being influenced by hits from other tracks or background noise. In this study, simulated single particle events are used to optimize the bin width. If at least one bin contains over 95% hits of the track, it is considered to be a good event. The proportion of good events to total events under different binning of the rho and alpha parameters was scanned. As an example, Fig. 6 shows the dependence of the efficiency of reconstructing good single muon events (three different momenta are shown) as functions of the bin width of rho and alpha. According to the scanning results, non-uniform bin widths varying with $\rho$ are used in the $\rho$ direction, and for the $\alpha$ direction, uniform binning with 800 bins spanning [0,π] is selected.

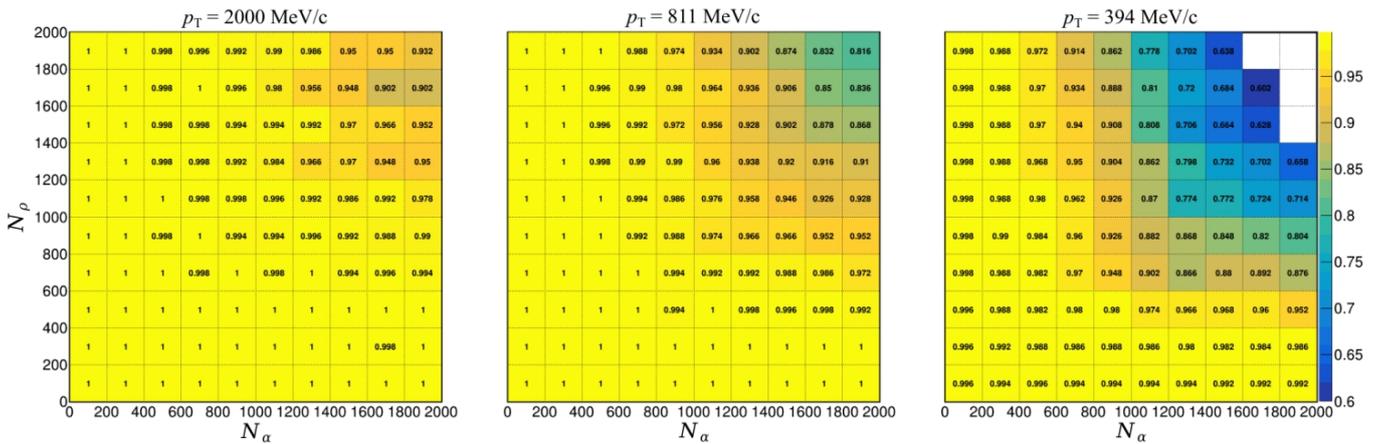

Fig. 6 Efficiency of good-event for simulated single muon with $p_z = 0$ and (left) $p_T = 2000$ MeV/c; (middle) $p_T = 811$ MeV/c; (right) $p_T = 394$ MeV/c under different bin widths. $N_\alpha$ and $N_\rho$ represent the number of bins in the $\alpha$ and $\rho$ directions with ranges of [0,π] and [-0.01,0.01], respectively. Uniformly divided bins are used.

**Track candidate estimation**

After filling the histogram, peaks will be identified by a specific peak finding method. Before searching for track candidates, a coarse-grained search is conducted to limit the range for finding peaks. Since the found peaks are local maxima, it is possible to have multiple peaks corresponding to the same track. Meanwhile, hits from different tracks or background may randomly form some peaks, these peaks are not desired to be retained. The identified peaks are filtered using certain optional cuts, such as the continuity of hits and the requirement for a minimum number of hits within a super-layer. Afterward, the parameters and shared hits of different peaks are compared to determine whether they should be merged. The retained peaks are considered as track candidates, with bin center ($\rho, \alpha$) as the initial parameters of the track. Afterwards, a least squares line fitting is performed on the conformal plane, followed by circle fitting using the Gauss-Newton method [9] to obtain more accurate circle parameters, with points too far from the fitted curve being discarded.

## 3.4 Stereo wire hits association and 3D track finding

At this stage, hits from the stereo layer of MDC are assigned to the found 2D track. If the projection of a stereo wire onto the XY plane intersects with the track circle, the $z$ position ($z_{rec}$) of this hit can be calculated. The accuracy of the calculated $z$-position is highly correlated with the accuracy of the estimated 2D track parameters. Although ITK hits are also assigned to the track in the 2D track finding, it is found that using only hits from the MDC for fitting the track parameters may lead to higher accuracy in calculating $z$, as particles are affected by material before entering the MDC. For each stereo layer hit, two $z$ position solutions can be obtained, and hits from other tracks may also be incorrectly assigned to the circular track, as shown in Fig. 7. Thus track finding in the $s$-$z$ space is also need, the projection of the track in the $s$-$z$ space is a straight line which can be described by equation

$$z_0 = z_{rec} - s_{rec} \cdot \tan \lambda,$$

where $s_{rec}$ is the path length from the POCA to the hit. Track finding in the $s$-$z$ plane is similar to two-dimensional track finding in the conformal plane, with a notable difference being that for one circle track, there should be only one matching $s$-$z$ trajectory. ITK hits with $z$ information and stereo-wire hits are used to determine track in parameter space. If a sufficient number of ITK hits have already been assigned to the track in the 2D track finding stage, the estimation of ($z_0, \tan\lambda$) can be carried out initially to constrain the search range when finding peaks.

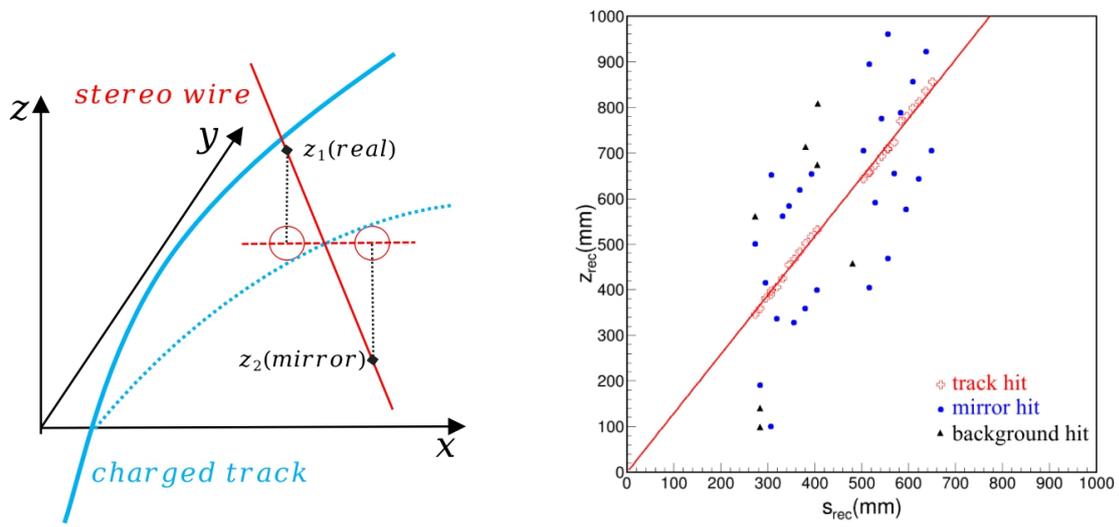

Fig. 7 Calculation of $z$ position (left), the blue dashed line and the red dashed line represent the projections of the track and the stereo wire in the XY plane, respectively. Each stereo wire measurement corresponds to two z-coordinates. Hits matched to a circular track in the SZ plane (right).

**3.5 Track fitting**

Finally, the deterministic annealing filter (DAF) provided by GENFIT2 [10] is used to do the track fitting, different particle assumptions are used to correct the materials effect of energy loss and scattering. After track fitting, the track will be extrapolated to the POCA to the origin, and the corresponding helix parameters will be obtained.

**4. Performance study**

**4.1 Background level**

The high luminosity of STCF introduces a significant background that cannot be overlooked in reconstruction. Background simulation is also implemented in OSCAR, hits generated from the background particles can be mixed with signal hits before reconstruction. Table. 2 lists the average number of background ITK hits and MDC hits mixed in an event. The numbers in the labels correspond to the ITK layer or MDC super-layer numbers.

Table. 2 Background hits count per event

| ITK1 | ITK2 | ITK3 | MDC1 | MDC2 | MDC3 | MDC4 | MDC5 | MDC6 | MDC7 | MDC8 |
|---|---|---|---|---|---|---|---|---|---|---|
| 37.3 | 13.6 | 8.2 | 60.3 | 42.4 | 24.8 | 25.1 | 60.0 | 67.8 | 30.8 | 30.0 |

**4.2 Definition of track finding efficiency**

Considering the acceptance of the detector and the interaction between particles and materials, only primary particles that produce sufficient hits to determine the trajectory parameters, are used for performance studies. The track finding efficiency presented represents the efficiency of the algorithm itself and does not account for the effect of the cases where particles fail to create enough hits due to detector acceptance or interactions. If a particle decays within the detectors, the hits produced by the secondary particles are considered as background. The information of simulated particle with corresponding hits is recorded as MC-track. The hit purity is the fraction of hits that belong to a particular MC-track of a reconstructed track. Hit purity represents the proportion of hits belonging to a particular MC-track within a reconstructed track, and a reconstructed track is considered to match with the MC-track if its hit purity exceeds 0.8. Track finding efficiency is defined as the fraction of particles which have at least one matched reconstructed track.

**4.3 Performance in reconstructing single particle events**

This section presents the reconstruction performance for single particles. Fig. 8 illustrates the track finding efficiency for e and μ. Fig. 9 shows the track finding efficiency for μ with momenta from 200-300 MeV, as a function of $cos\theta$. When the particle's transverse momentum is very low, it cannot pass through enough detector layers, leading to a decrease in tracking efficiency. Additionally, the track parameters undergo significant changes due to the particle's energy loss and multiple scattering, making it difficult to determine the complete particle trajectory and its parameters, which results in track finding failure, especially at low momenta. At high momenta, the track finding efficiency for electrons is slightly lower than that of muons due to the effects of bremsstrahlung. Reconstruction of particles with low momentum and large polar angles is more strongly affected by background. Fig. 10 shows the resolution of $p$ for μ, the resolution is obtained by fitting the residuals with a Gaussian function. The resolution for single particles with background is nearly the same compare to the case without background, thus is not shown. At a momentum of 1 GeV, the relative momentum resolution is better than 0.5% when the particle exits through the barrel of the MDC. As the value of $cos\theta$ increases, the momentum resolution slightly decreases due to the stronger influence of material effects. However, when the particle exits through the endcap (with a large dip angle), the momentum resolution rapidly worsens as the number of detector layers it passes through decreases.

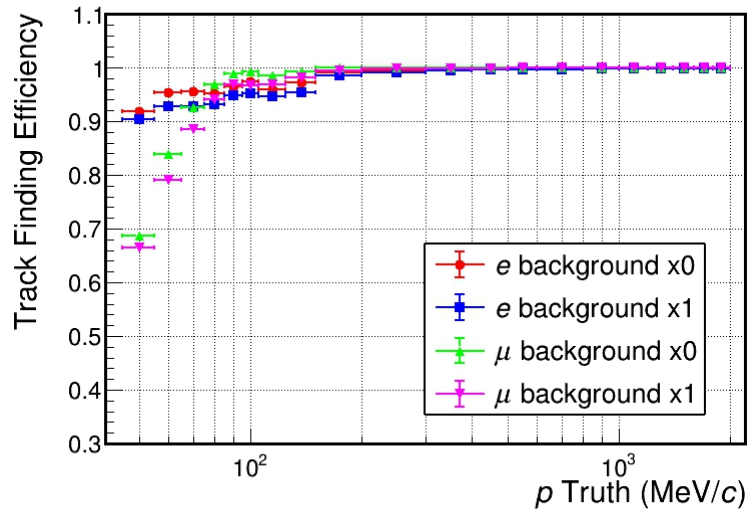

Fig. 8 The track finding efficiency for single e and μ as functions of $p$, with $|cos\theta|<0.8$.

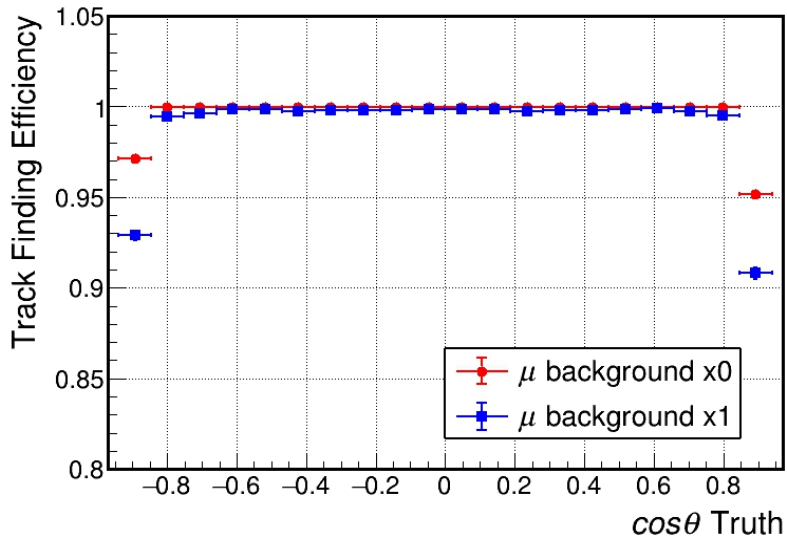

Fig. 9 The track finding efficiency for single μ with $p$ ranges of [200,300] MeV, as functions of $cos\theta$.

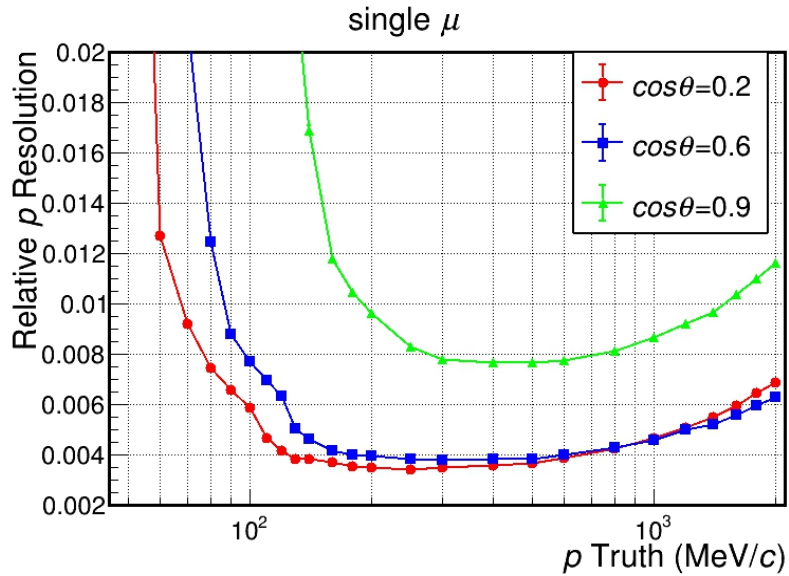

Fig. 10 The resolution of $p$ for μ without background.

**4.4 Performance in reconstructing ψ(3686) → π+π- J/ψ, J/ψ → μ+μ- events**

The simulated ψ(3686) → $\pi^+\pi^-$ J/ψ, J/ψ → $\mu^+\mu^-$ events are used to evaluate the track reconstruction performance. Fig. 11 shows the distributions of $p_T$ versus $cos\theta$ for μ and π in the ψ(3686) → $\pi^+\pi^-$ J/ψ events. The transverse momentum of particles is correlated with its polar angle. Generally speaking, the smaller the transverse momentum, the larger the dip angle.

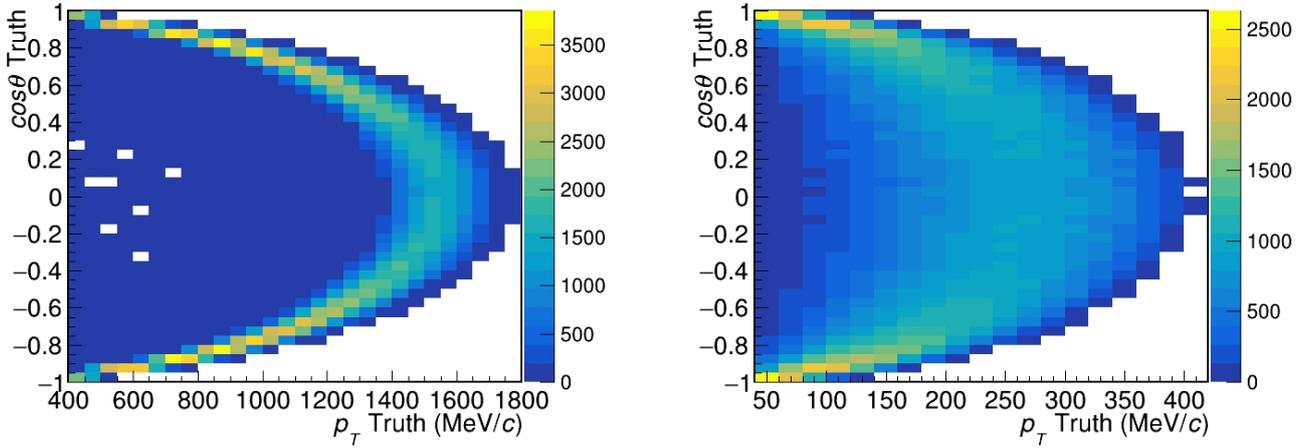

Fig. 11 The distributions of $p_T$ versus $cos\theta$ for μ (left) and π (right) in ψ(3686) → π+π- J/ψ, J/ψ → μ+μ- events

Fig. 12 shows the track finding efficiency for different particles in the ψ(3686) → π⁺π⁻ J/ψ, J/ψ → μ⁺μ⁻ events. Since the detection efficiency of the detector is unlikely to reach 100%, a portion of the hits are randomly excluded during the track reconstruction phase to take this effect into account. As shown in Fig.12, even if the detection efficiency is reduced to 90%, a high track finding efficiency is still achieved. Fig. 13 shows the resolution of the helix parameters $d_0$, $z_0$ and $p$. The resolution slightly decreases because the reduction in detection efficiency result in fewer points being available for track fitting. Because the particle's transverse momentum is correlated with its polar angle, for μ, when the transverse momentum decreases and the number of detector layers the particle passes through becomes too few, the track finding efficiency and resolutions decrease significantly. For π, as the transverse momentum decreases, not only does the number of detector layers passed through reduce, but the effects of material interactions also become very significant, which have a negative impact on the reconstruction performance. Fig.14 shows the track finding efficiency for different background level. The track finding efficiency decreases due to the impact of the background hits, while the resolution of the parameters remains nearly unchanged. As the transverse momentum decreases and the dip angle increases, the reconstruction performance becomes more susceptible to the impact of background noise.

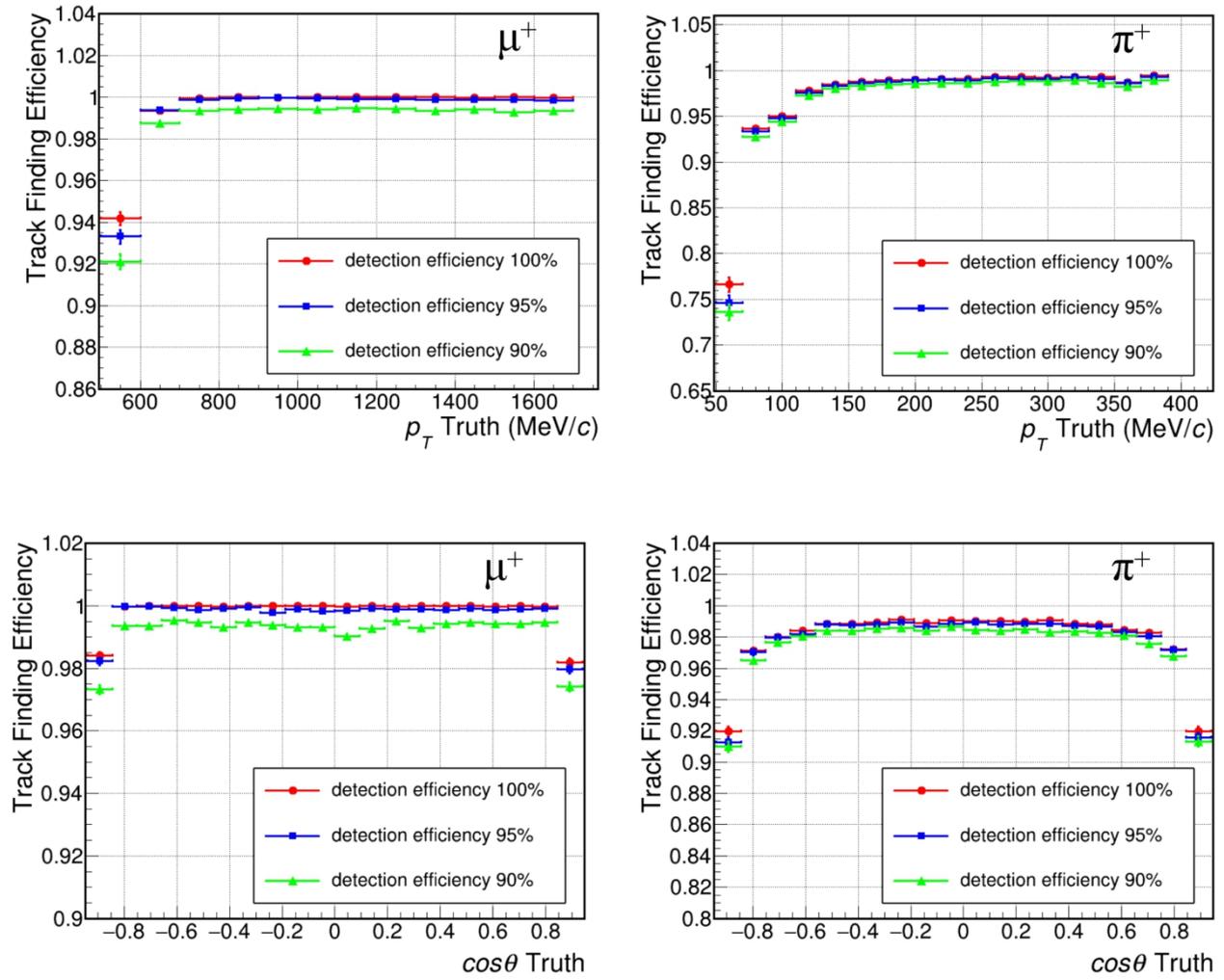

Fig. 12 The track finding efficiency as functions of $p_T$ and $cos\theta$, for $\mu^+$(left) and $\pi^+$(right) in $\psi(3686) \to \pi^+\pi^-$ J/$\psi$, J/$\psi \to \mu^+\mu^-$ events.

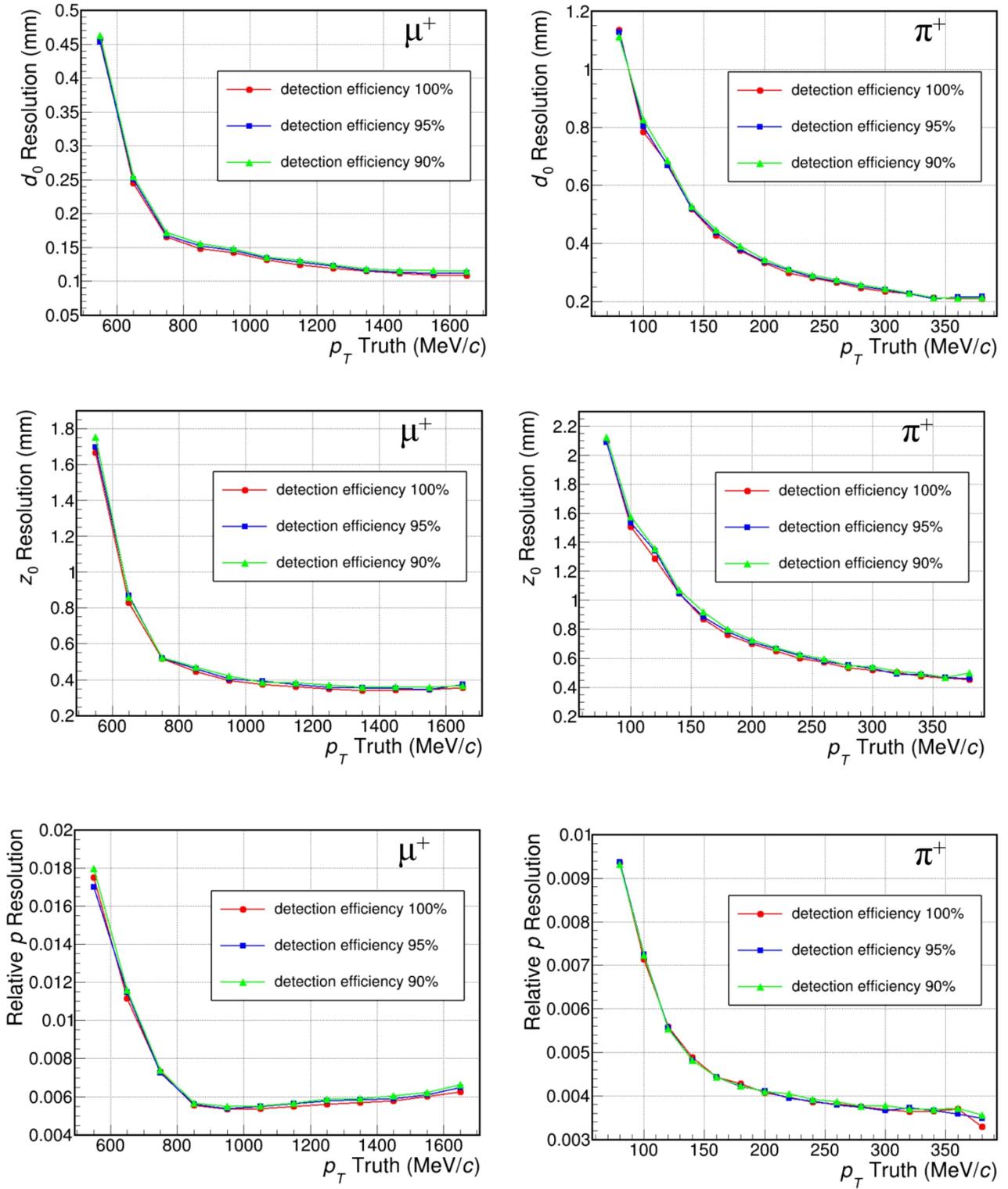

Fig. 13 The resolution of $d_0$, $z_0$ and $p$ for μ⁺(left) and π⁺(right) in ψ(3686) → π+π- J/ψ, J/ψ → μ+μ- events.

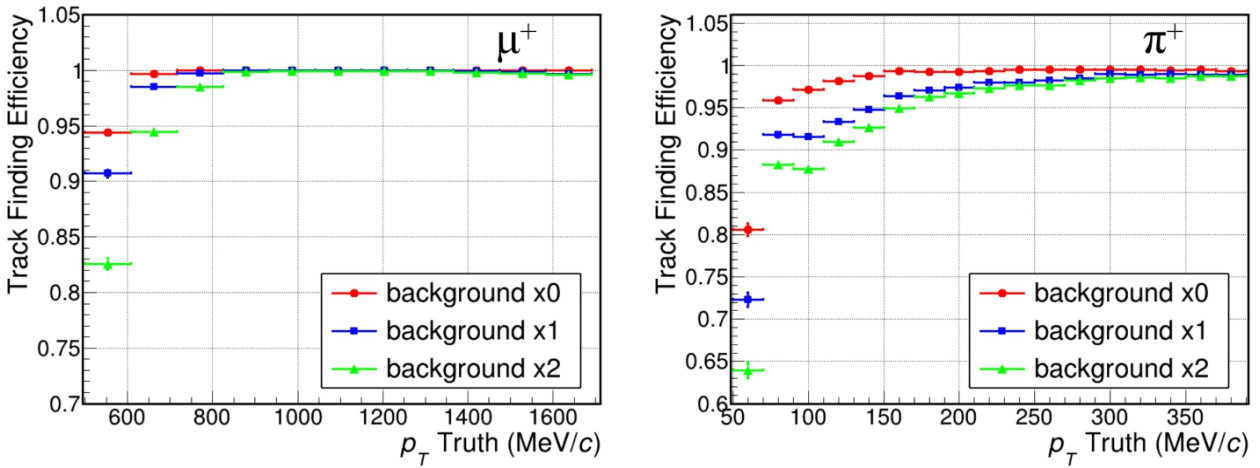

Fig. 14 The track finding efficiency with background and without background as a function of $p_T$, for μ+(left) and π+(right) in ψ(3686) → π+π- J/ψ, J/ψ → μ+μ- events. Assuming 100% detection efficiency.

## 5. Conclusion

A global track finding algorithm based on Hough transform is implemented in STCF offline software framework. This method handles different types of measurements from different detectors. Based on the detector design of a 3-layer of ITK uRWELL and a 48-layer MDC, the performance study, including scenarios with reduced detector efficiency and different background levels, shows that a high tracking efficiency can be achieved to meet the requirements of STCF. As a global algorithm, it demonstrates robustness against local detector inefficiencies. For particles exiting from the barrel of the MDC, the relative momentum resolution is better than 0.6% in most of momentum ranges. Further research and optimization are needed for this algorithm in more sophisticated circumstances, such as when dealing with secondary vertex particle tracks and even higher background level environments. In order to better fulfill the physical goals of STCF, other methods for track reconstruction, such as the utilization of the combinatorial kalman filter(CKF) [11], as well as background filtering and track finding based on machine learning, are also under research.

**Acknowledgement**

We thank the University of Science and Technology of China and Hefei Comprehensive National Science Center for their strong support on the promotion and development of the STCF project. This work was supported by the National Natural Science Foundation of China (Grant No. 12125505 and No.12375197) and the STCF key technology research and development project.